\documentclass[aps,a4paper,10pt]{revtex4}

\usepackage{graphicx}
\usepackage{hyperref}

\usepackage{subfigure}
\usepackage{amstext}
\usepackage{amssymb}
\usepackage{amsmath}
\usepackage{float}
\usepackage{wrapfig}

\newcommand{\om}{\omega}

\newcommand{\bra}{\langle}
\newcommand{\ket}{\rangle}

\newcommand{\non}{\nonumber}
\newcommand{\be}{\begin{equation}}
\newcommand{\ee}{\end{equation}}

\begin{document}

\DeclareGraphicsExtensions{.eps, .jpg}


\title{
       {\parbox[b]{\textwidth}{\rm \footnotesize \begin{flushright}
       K. H. Hughes (ed.) \\
       Dynamics of Open Quantum Systems \\
       \copyright \hspace{1ex} 2006, CCP6, Daresbury
        \end{flushright}}} \\
Time-convolutionless master equation dynamics for charge-transfer processes 
between semiconducting polymers}
\author{Eric R. Bittner and Andrey Pererverzev}
\affiliation{Department of Chemistry, University of Houston, Houston TX 77204}


\maketitle
\setcounter{page}{1}
\thispagestyle{plain}

The dynamics of charge transfer between molecular species in the condensed phase
remains a difficult and compelling problem from a theoretical perspective.  
The difficulty stems from the fact that one is faced with the situation that for most of the time
there is a clear separation in the time-scales between the motion of the electronic 
degrees of freedom and the nuclear or molecular degrees of freedom.   
Hence, most of the time, nuclear reorganization dynamics following some 
dramatic change in the electronic structure occurs on the Born-Oppenheimer 
potential energy curves parameterized by the nuclear coordinates.   However, 
this description  breaks down whenever the electronic wave function 
changes rapidly in the direction of the nuclear motion leading to transitions between
the discrete electronic states.   This coupling becomes even more dramatic when there
is an actual crossing of the potential curves at certain nuclear geometries.   
Even more so, the number of nuclear degrees of freedom, while finite, 
 typically far out number  the number of relevant electronic states for a given process.  

For the case of a infinite thermostat, the Pauli master equation and Redfield equation
have long been applied to the study of quantum relaxation.    While originally 
derived from more or less heuristic arguments, such equations can be obtain 
from formally exact master equations using projection operator techniques 
and a series of well defined approximations.    In the 
limit of a continuum of modes, the resulting transition probabilities become identical 
to what one obtains using Fermi's golden rule.  

In this paper, we review our recent work in deriving a formally exact and time-local 
approach for incorporating non-Markovian dynamics in to the quantum master equation 
for state to state electronic transitions. \cite{pereverzev:104906}  Here, we discuss some of the
details of this work with particular attention to its theoretical development. 
As an application, we  examine the kinetics of charge transfer between two co-facially stacked 
conjugated polymer chains \cite{morteani:244906,morteani:1708,morteani:247402}
using a  model recently developed by our group. \cite{bittner:214719}
Here we compare the inclusion of both fast and slow phonon modes affect the 
transfer of a hole from one chain to the next following photo-excitation. 

The Hamiltonian describing a wide range of photphysical problems can be cast in the form:
\begin{eqnarray} H=\sum_n\epsilon_n |n\ket\bra
n|+\sum_{nmi} g_{nmi}|n\ket\bra m|(a^{\dagger}_i+a_i)
+\sum_i\om_ia^{\dagger}_ia_i. \label{Ham} 
\end{eqnarray} 
Here $|n\ket$'s denote
electronic states with vertical energies $\epsilon_n$, $a_i^{\dagger}$
and $a_i$ are the creation and annihilation operators for the normal
mode $i$ with frequency $\omega_i$, and $g_{nmi}$ are the coupling
parameters of the electron-phonon interaction which we take to be
linear in the phonon normal mode displacement coordinate.

We can separate $H$ into a part that is diagonal with respect to the
electronic degrees of freedom,
\begin{eqnarray} H_0=\sum_n\epsilon_n |n\ket\bra n| +\sum_{ni}g_{nni}|n\ket\bra
n|(a^{\dagger}_i+a_i) +\sum_i\om_ia^{\dagger}_ia_i, \label{H0}
\end{eqnarray} 
and an off-diagonal part $V$ 
\begin{eqnarray} V={\sum_{nmi}}'g_{nmi}|n\ket\bra
m|(a^{\dagger}_i+a_i), \label{V} 
\end{eqnarray}
 where the prime at the summation sign indicates that the terms with
$n=m$ are excluded.  This separation is useful for the following two
reasons. First, in many systems only off-diagonal coefficients
$g_{nmi}$ are small compared to $g_{nni}$.
Hence, $V$ can be treated as a perturbation. Second, for many cases of
interest, the initial density matrix commutes with $H_0$. In this
case, the separation gives simpler forms of the master equations.

For further analysis it is convenient to perform a polaron transform using
\begin{eqnarray}
U=e^{-\sum_{ni}\!\!\frac{g_{nni}}{\om_i}|n\ket\bra
n|(a^{\dagger}_i-a_i)}= \sum_{n}|n\ket\bra
n|e^{-\sum_{i}\!\!\frac{g_{nni}}{\om_i}(a^{\dagger}_i-a_i)}
\label{unitary} 
\end{eqnarray} 
in which our transformed Hamiltonian becomes
\begin{eqnarray} \tilde H_0=U^{-1}H_0U
=\sum_n\tilde\epsilon_n |n\ket\bra n|+\sum_i\om_ia^{\dagger}_ia_i,
 \end{eqnarray}
where the renormalized electronic energies are 
\begin{eqnarray}
\tilde\epsilon_n=\epsilon_n-\sum_{i}\frac{g_{nni}^2}{\omega_i}.  
\end{eqnarray}
Applying the same unitary transformation to $V$ gives 
\begin{eqnarray}
\tilde{V}=\sum_{nmi} |n\ket\bra m|M_{nmi}.
 \label{opm}, 
\end{eqnarray}
where the system-bath operators are
\begin{eqnarray} M_{nmi}=g_{nmi}\left(a^{\dagger}_i+
a_i-\frac{2g_{nni}}{\omega_i}\right)e^{\sum_{j}\frac{(g_{nnj}-g_{mmj})}{\om_j}(a^{\dagger}_j-a_j)}
\label{opm}.
\end{eqnarray}
At this point it is useful to connect the various terms in our Hamiltonian 
with specific physical parameters. 
The terms involving $(g_{nnj}-g_{mmj})/\omega_j$ 
can be related to the reorganization energy
$$
E^\lambda_{nm} =\sum_j \frac{(g_{nnj}-g_{mmj})^2}{\omega_j^2} = \sum_j \hbar\omega_j S_j
$$ where $S_j$ is the Huang-Rhys factor for mode $j$ which is related
to the Franck-Condon factor describing the overlap between the $v_j=1$
vibronic state in one electronic state with the $v_j=0$ vibronic state
in the other.  Likewise, the energy difference between the
renormalized energies is related to the driving force of the
transition,
$$
\Delta E_{nm} = \tilde \epsilon_n-\tilde \epsilon_m.
$$ In the transformed picture the electronic transitions from state
$|n\ket$ to $|m\ket$ are accompanied not only by the creation or
annihilation of a single phonon of mode $i$ but also by the
displacements of all the normal modes.  This is quite different from
the spin-boson model which does not have coordinate dependent coupling
between the electronic states.

For a properly chosen projection operator ${\cal P}$ and the initial
total density matrix that satisfies $\rho(0)={\cal P}\rho(0)$, ${\cal
P}\rho(t)$ can be shown to satisfy at least two different master
equations: the Nakajima-Zwanzig (NZ) equation \cite{Nakajima, Zwanzig,
Prigogine} and the time-convolutionless (CL) master equation
\cite{Shibata}.
\begin{eqnarray}
\frac{\partial {\cal P}\rho(t)}{\partial t}&=&
-\int_0^td\tau {\cal K}^{NZ}(t-\tau) {\cal P}\rho(\tau), \label{NZ}\\
\frac{\partial {\cal P}\rho(t)}{\partial t}&=&
-\int_0^td\tau{\cal K}^{CL}(\tau) {\cal P}\rho(t)\label{CL}.
\end{eqnarray}
The explicit expressions for superoperators ${\cal K}^{NZ}(\tau)$ and
${\cal K}^{CL}(\tau)$ can be found in Ref. \cite{Breuer}. Since
Eq. \ref{CL} is less well known than Eq. \ref{NZ} we will show here
how Eq. \ref{CL} can be derived from Eq. \ref{NZ}. Applying the
Laplace transformation to both sides of Eq. \ref{NZ} we obtain \be
s{\cal P}\rho(s)-{\cal P}\rho(0)=-{\cal K}(s){\cal P}\rho(s), \ee
where $\rho(s)$ and ${\cal K}(s)$ are the Laplace transforms of
$\rho(t)$ and ${\cal K}(t)$, respectively.  Solving the last equation
for $\rho(s)$ and applying the inverse Laplace transformation we have
the formal solution of Eq. \ref{NZ} \be {\cal P}\rho(t)={\cal
V}(t){\cal P}\rho(0),\label{prop} \ee where \be {\cal
V}(t)=\frac{1}{2\pi
i}\int_{\kappa-i\infty}^{\kappa+i\infty}\frac{e^{st}}{s+{\cal K}(s)}
\ee and $\kappa$ is an arbitrary positive constant chosen so that the
contour of integration lies to the right of all singularities of the
integrand. Differentiating Eq. \ref{prop} with respect to time we
obtain the time local master equation \be \frac{\partial {\cal
P}\rho(t)}{\partial t}= {\cal F}(t) {\cal P}\rho(t) \label{timeloc}
\ee with \be {\cal F}(t)=\frac{\partial {\cal V}(t)}{\partial t}{\cal
V}^{-1}(t).  \ee Eq. \ref{timeloc} can be written in the form of
Eq. \ref{CL} with \be {\cal K}^{CL}(\tau)=-\frac{\partial{\cal
F}(\tau)}{\partial \tau}.  \ee

Even though  Eqs. \ref{NZ} and \ref{CL} are formally exact,  it is 
impossible to determine ${\cal K}^{NZ}(\tau)$ and ${\cal K}^{CL}(\tau)$ for most realistic systems.
Since we have assumed the coupling to be weak, we 
ond order in the coupling constants 
\begin{eqnarray}
{\cal K}_2^{NZ}(\tau)={\cal K}_2^{CL}(\tau)=
{\cal PL}_Ve^{-i{\cal L}_0\tau}{\cal L}_V{\cal P}. \label{comparison}
\end{eqnarray}
Here ${\cal L}_0$ and ${\cal L}_V$ are the Liouville superoperators corresponding
to $\tilde H_0$ and $\tilde V$ whose action on some 
 density matrix $\rho$ is given by
\begin{eqnarray}
{\cal L}_0\rho=\tilde H_0\rho-\rho\tilde H_0, \qquad 
{\cal L}_V\rho=\tilde V\rho-\rho\tilde V.
\end{eqnarray}
In Ref.~\cite{Breuer}, Breuer and Petruccione show that to second
order in the coupling constants, the convolutionless experession
(Eq. \ref{CL}) gives a better approximation to the exact solution than
the Nakajima-Zwanzig equation (Eq. \ref{NZ}).  It also has an
additional mathematical convenience of being local in time. Therefore,
in the following analysis we will use the convolutionless approach.

Taking our initial density matrix in the transformed representation as 
\begin{eqnarray}
\tilde\rho(0)=|n\ket\bra n|
\frac{e^{-\beta{\sum_i\om_ia^{\dagger}_ia_i}}}
{{\rm Tr}(e^{-\beta{\sum_i\om_ia^{\dagger}_ia_i}})}. \label{transrho}
\end{eqnarray}
we will use the
projection operator that acts on the total density matrix in the
following way
\begin{eqnarray}
{\cal P}\rho=\sum_n|n\ket\bra n|\rho_{eq}^{os}\,
{\rm Tr}\left(|n\ket\bra n|\rho\right), \label{P}
\end{eqnarray}
where 
the trace is taken over both electronic and oscillator degrees of freedom and
\begin{eqnarray}
\rho_{eq}^{os}=\frac{e^{-\beta{\sum_i\om_ia^{\dagger}_ia_i}}} {{\rm
Tr}(e^{-\beta{\sum_i\om_ia^{\dagger}_ia_i}})}. \label{equli}
\end{eqnarray}
Note that 
\be
{\rm Tr}\left(|n\ket\bra n|\rho\right)=P_n,
\ee
where $P_n$ is the probability to
find the electronic subsystem in state $|n\ket$.
Using Eq. \ref{CL}, the definition of ${\cal K}_2^{CL}(\tau)$ in Eq. \ref{comparison}, and the definition of
${\cal P} \rho$ given by Eq. \ref{P} the following explicit 
convolutionless equation is obtained 
\begin{eqnarray}
\frac{d P_n}{d t}=\sum_mW_{nm}(t)P_m-\sum_mW_{mn}(t)P_{n} \label{Pauli}.
\end{eqnarray}
The time dependent rates $W_{nm}(t)$ are  given by
\begin{eqnarray}
W_{mn}(t)=2\Re e\int_0^td\tau \sum_{ij}\langle M_{nmi}M_{mnj}(\tau)\rangle
e^{-i(\tilde\epsilon_n-\tilde\epsilon_m)\tau}, \label{rates}
\end{eqnarray}
where 
\begin{eqnarray}
\langle M_{nmi}M_{mnj}(\tau)\rangle=
{\rm Tr}\left(M_{nmi}M_{mnj}(\tau)\rho^{os}_{eq}\right). \label{corrfun}
\end{eqnarray} 
Due to the explicit form of operators $M_{nmi}$ (Eq. \ref{opm}) the
calculation of the correlation functions in Eq. \ref{corrfun} can be
reduced to the averaging of the displacement operators over the
equilibrium ensemble (Eq. \ref{equli}). After straightforward, but
lengthy, calculations we obtain the principal result of our work:
\begin{eqnarray}
\langle M_{nmi}M_{mnj}(\tau)\rangle&=&g_{nmi}g_{mnj} \non \\
& \times&\left(\left(\Delta_{nmi}(\overline{n}_i+1)e^{i\om_i\tau}
-\Delta_{nmi}\overline{n}_ie^{-i\om_i\tau}+\Omega_{nmi}\right) \right.\non \\
 &\times&\left.\left(\Delta_{nmj}(\overline{n}_j+1)e^{i\om_j\tau}
-\Delta_{nmj}\overline{n}_je^{-i\om_j\tau}+\Omega_{nmj}\right)\right.\non \\
&+ &\left.\delta_{ij}(\overline{n}_i+1)e^{i\omega_i\tau}+\delta_{ij}\overline{n}_i
e^{-i\omega_i\tau}\right)q_{nm}(\tau)f_{nm}(\tau).
 \label{corrf}
\end{eqnarray}
Here 
\begin{eqnarray}
\Delta_{nmi}&=&\frac{(g_{nni}-g_{mmi})}{\om_i}, \\
\Omega_{nmi}&=&\frac{(g_{nni}+g_{mmi})}{\om_i}, \\
\overline{n}_i&=&\frac{1}{e^{\beta\om_i}-1}, \\
q_{nm}(\tau)&=&e^{i\sum_j{\Delta^2_{nmj}\sin\omega_j\tau}},\label{q} \\ 
f_{nm}(\tau)&=&e^{-2\sum_j(\overline{n}_j+\frac{1}{2})\Delta_{nmj}^2(1-\cos\om_j\tau)}. \label{f}
\end{eqnarray}
In the case when all the diagonal electron/phonon terms vanish,
 $g_{nni}=0$, the correlation functions in Eq. \ref{corrf} reduce to
 those obtained within the golden rule approach \cite{May,Nitzan}. It is
 clear from Eq. \ref{corrf} that for systems with $g_{nni}/\omega_i
 \gg 1$
 the golden rule approach is not applicable.  
 Unfortunately,  the complicated form of the correlation functions 
 precludes much further analysis at this point and we need to turn to 
 numerical calculations.

\begin{figure}[t]
\includegraphics[width=0.74\columnwidth]{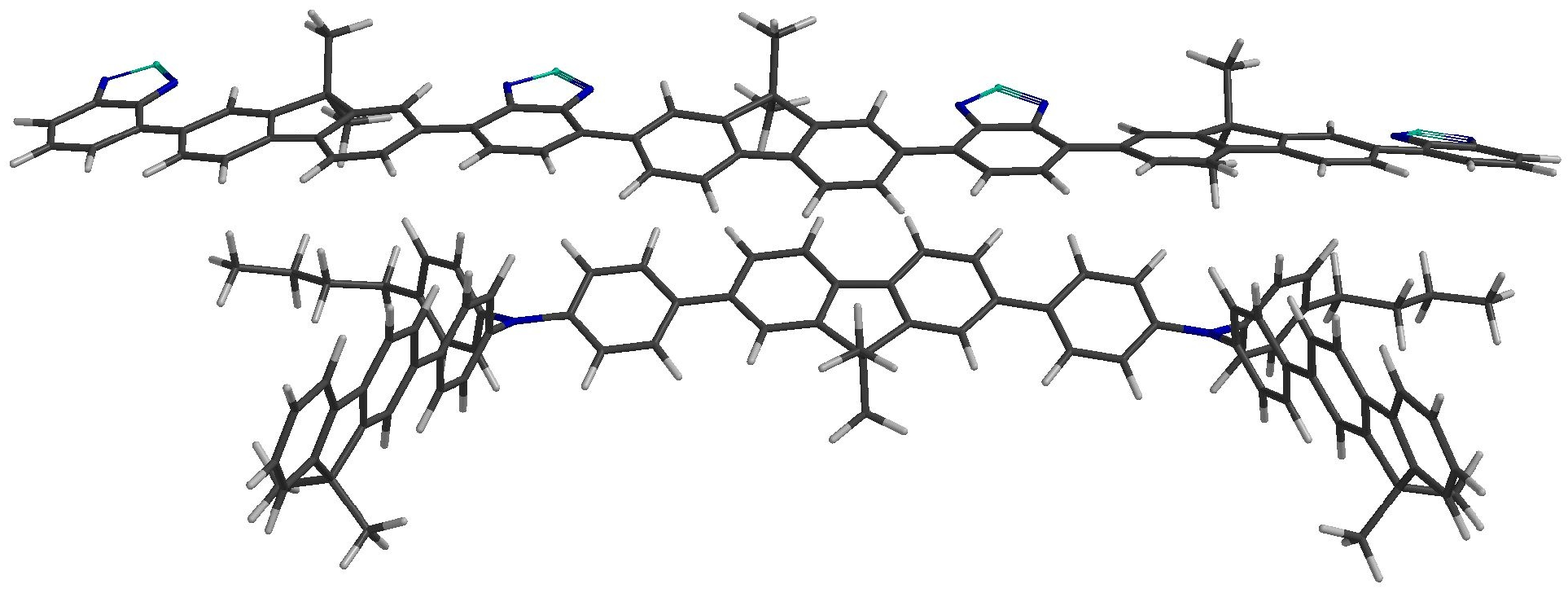}
\includegraphics[width=0.25\columnwidth]{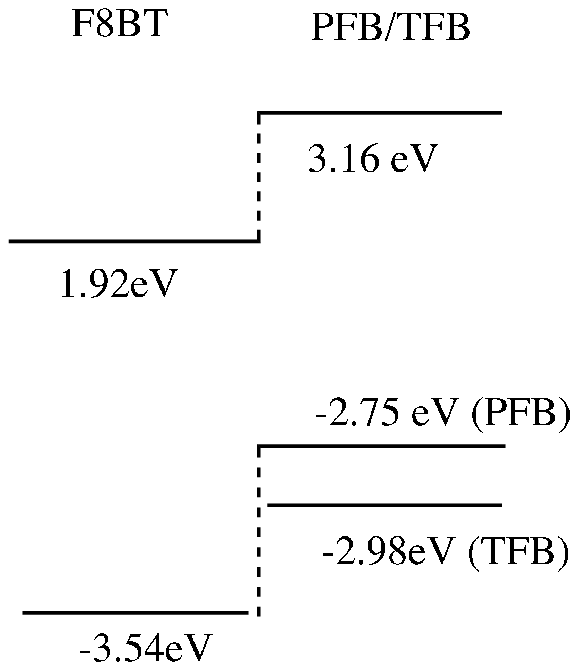}
\caption{(Left) Molecular geometry of TFB:F8BT interface.  (Right)
Relative energies of highest occupied and lowest unoccupied 
orbitals  of TFB and F8BT. }\label{stack}
\end{figure}
 
As an application of our approach, we consider the electronic relaxation in the
conjugated polymer heterojunctions previously investigated in
Ref. \cite{Ramon1, Ramon:2006} by a different approach.  
Here we consider the decay of an excitonic state into a charge-separated
state at the interface between two semiconducting polymer phases (TFB) and  
(F8BT).   The structure of the polymers at the interface is shown in Fig.~\ref{stack}.
 Such materials have been extensively studied for their potential in 
organic light-emitting diodes and organic photovoltaics
\cite{morteani:247402,russell:2204,silva:125211}.
 At the phase boundary, the material forms a type-II semiconductor
heterojunction with the off-set between the valence bands of the two
materials being only slightly more  than the binding energy of 
an exciton placed on either the TFB or F8BT polymer.  As a result, an exciton on 
the F8BT side will dissociate to form a charge-separated (exciplex) state at the 
interface. \cite{morteani:244906,morteani:1708,morteani:247402,silva:125211,stevens:165213}
$$
{\rm F8BT}^*:{\rm TFB}\longrightarrow {\rm F8BT}^-:{\rm TFB}^+.
$$
Ordinarily, such type II systems are best suited for photovoltaic rather than LED 
applications.
However, LEDs fabricated from phase-segregated 50:50 blends of TFB:F8BT 
give remarkably efficient electroluminescence efficiency due to {\em secondary} 
exciton formation in the back-reaction
$$
{\rm F8BT}^-:{\rm TFB}^+\longrightarrow {\rm F8BT}^*:{\rm TFB},
$$ 
as thermal equilbrium between the excitonic and charge-transfer
states is established.  This is evidenced by long-time emission,
blue-shifted relative to the emission from the exciplex, accounting
for nearly 90\% of the integrated photo-emission.

We consider only the two lowest electronic levels corresponding to 
$$
|XT\rangle =   {\rm F8BT}^*:{\rm TFB} \,\,\&\,\, |CT\rangle= {\rm F8BT}^-:{\rm TFB}^+.
$$ 
We take the vertical energies for these two states as $\epsilon_{CT} = 2.191$eV and
$\epsilon_{XT}= 2.294$ eV. 
As was shown in Ref. \cite{khan:085201, Bittner1, Bittner2, Bittner3}, in 
poly-fluorene based systems, 
there are essentially two groups of phonon modes that are coupled strongly
to the electronic degrees of freedom as evidenced by their presence as
vibronic features in the vibronic emission spectra, namely: low
frequency torsional modes with frequencies between 90 and 100 cm$^{-1}$
and higher frequency C=C stretching modes with frequencies
 between 1500 and 1600 cm$^{-1}$ \cite{khan:085201}.
The vibronic couplings within the model were determined by 
comparison between the Franck-Condon peaks of the predicted and observed spectra of the 
system.

\begin{figure}
 \includegraphics[width=0.4\columnwidth]{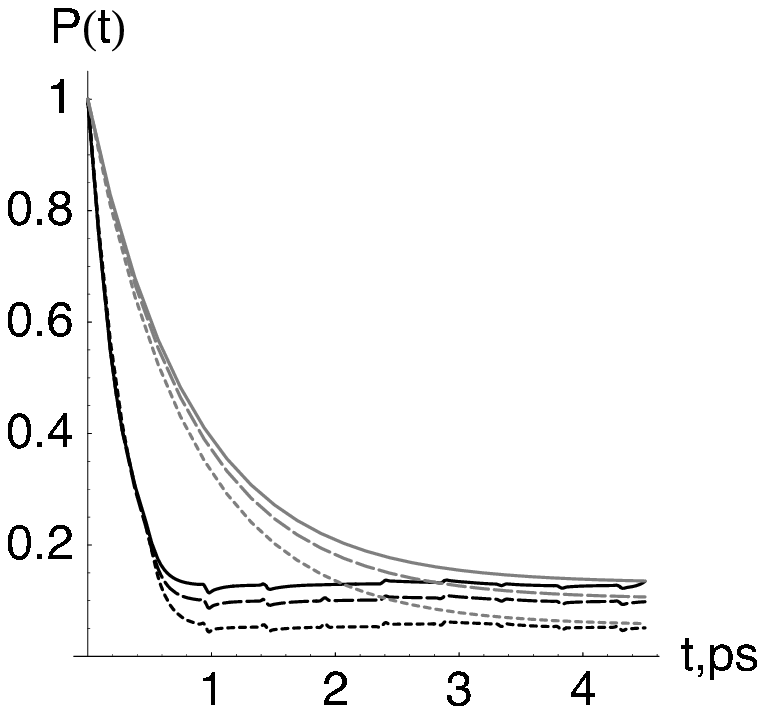}  
  \includegraphics[width=0.4\columnwidth]{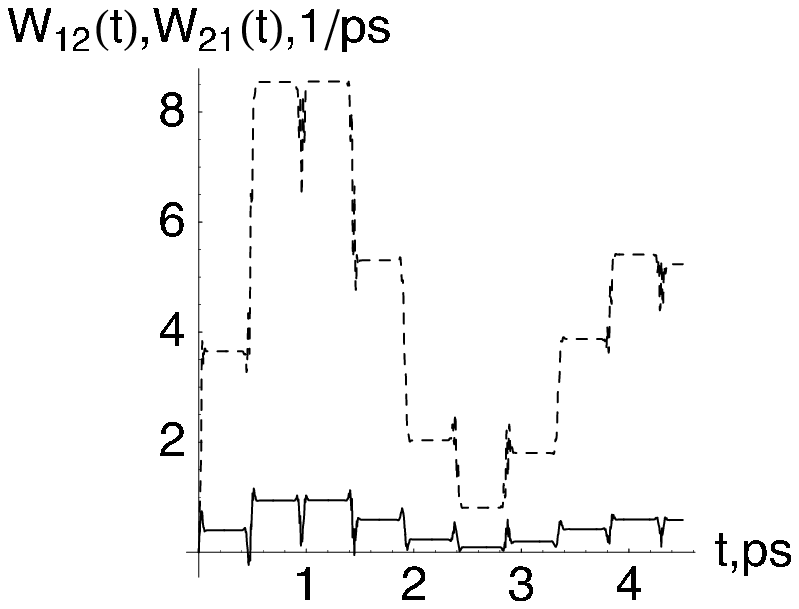}  
 \caption{(Left) The population of the excited state as a function of
 time for the TFB:F8BT heterojunction for three different temperatures
 obtained by solving the time-convolutionless master equation (black)
 or through the Marcus-type approximation (gray) at $T =$ 230K, 298K,
 and 340K.  (Right) Transition rates $W_{12}(t)$ and $W_{21}(t)$ for
 the TFB:F8BT heterojunction involving 24 normal modes at $298$
 K.}\label{Figure3}
 \end{figure}

As before, we consider the initial state as being prepared in the
$|XT\rangle$ state corresponding to photoexcitation of the F8BT
polymer.  The results for the higher electronic level population
obtained by numerically solving Eq. (\ref{Pauli}) are shown on
Fig. \ref{Figure3}.  The time dependence of coefficients $W_{12}(t)$
and $W_{21}(t)$ is shown on Fig. \ref{Figure3}.  As in the previous
example the electronic relaxation does not follow a simple exponential
pattern.  The main difference between this model and the previous
example is that the electronic relaxation time is of the same order as
the recurrence time for coefficients $W_{12}(t)$ and $W_{21}(t)$. As
can be seen from Fig. \ref{Figure3} the time dependence of these
coefficients has the form of approximately constant regions abruptly
changed at regular recurrence intervals. This type of time dependence
makes the electronic relaxation look like a series of exponential
relaxations with changing rates. We do note, however, that  
transition rates $W_{12}(t)$ and $W_{21}(t)$ can become negative as the 
relaxation proceeds towards the equilibrium population.  As with any approximate
method, this can lead to a lack of positivity in the populations.

Note also, that  because the relaxation does not obey a simple 
exponential rate law, the initial decay slightly ``overshoots'' the final 
equilibrium population.  This is most evident in the highest temperature 
case considered here (T = 340K).   Since the $|XT\rangle$ is also  the emissive 
state, photo-emission (not included herein) depletes this population on a
nanosecond time scale (radiative lifetime).  
Population  back-transfer from the $|CT\rangle$ to maintain a thermal equilibrium population
then leads to the continuous replenishment of the emissive species
such that nearly all of the CT population contributes to the formation of 
secondary (regenerated) excitonic states
\cite{morteani:247402,russell:2204,silva:125211,bittner:214719,Ramon:2006}.

\begin{figure}[t]
\includegraphics[width=0.5\columnwidth]{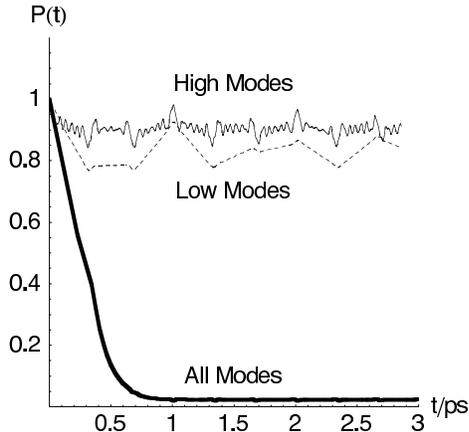}
\caption{Exciton population keeping only the high frequency (solid), low frequency, or using 
both high and low frequency phonon modes.}\label{modes}
\end{figure}

It is of interest to compare the relaxation dynamics in TFB:F8BT
heterojunction obtained through the application of the
time-convolutionless master equation with other approaches. As an
example we will consider the Marcus-type approach which is widely used
to study electron transfer in chemical systems \cite{May, Ramon:2006,
Nitzan:2006, Stuch}.     Based upon the model parameters, the driving force and 
reorganization energy are 
$\Delta E_{21} = 0.073$ eV  and $E_{\lambda} = 0.092 $ eV, which places the 
system very close to the barrier-less regime as reported by 
Morteani, {\em et al}~\cite{morteani:1708}. 
Because of the coordinate dependency of the coupling, some care must be 
taken in deriving the Marcus rates and we give details of this in the 
appendix.
The results for the relaxation dynamics using these
rates applied to TFB:F8BT heterojunction for three different
temperatures are shown in Fig. \ref{Figure3}.  

It can be seen that
apart from a more complicated time dependence the master equation
approach gives faster relaxation when compared to the Marcus-type
picture. The discrepancy between the two approaches can be explained
by the fact that the Marcus approximation is assumed to be valid when
$kT \gg \hbar \omega_i$ for all the normal modes. In the case of the
TFB:F8BT heterojunction this condition is not satisfied for the higher
frequency modes.

Finally, we consider the contributions of each type of intramolecular motions 
to the charge-transfer process.  Recall, our model included two non-overlapping bands
of intramolecular phonon modes:  a low frequency band corresponding to the 
torsional motions of the chain and a  high frequency band corresponding to C=C bond
stretching motions.   In Fig.~\ref{modes} we show the contribution of each band to the 
overall charge-transfer process.  If we include only one band or the other, {\em very little
charge transfer occurs} while if we include both bands, transfer occurs within the first 0.5ps
following excitation.    This emphasizes the fact that while one set of modes may be 
bringing the system into and out of regions of strong coupling between the XT and CT 
states (likely the high frequency modes), the other set of modes are required to 
help dissipate the energy associated with making the transition.

In conclusion, we present here a brief overview of a new approach we have developed for
incorporating non-Markovian dynamics into the calculation of state-to-state transition 
probabilities for electronic systems.   The time evolution of the electronic populations is,
in general, more complex then the one obtained with the Pauli master
equation and depends on the explicit form of the time dependent
coefficients.  The time-dependence of the rate coefficients introduces non-Markovian
effects due to the vibrational motion into the electronic population transfers.  
What is currently lacking, however, is the inclusion of the electronic coherences and the 
important process of decoherence.  Such contributions are quite important and we are 
currently extending our approach in that direction.    
While more complete and fully quantum approaches (such as MCTDH)
permit highly detailed analysis of the dynamics of systems such as these, the computational 
cost associated with these approaches remains quite high.

\begin{acknowledgments}
This work was supported in part by grants from the National Science Foundation (US) and 
the Robert Welch Foundation.    We also than the organizers of the CCP6 for providing a 
beautiful place for a highly stimulating workshop.
\end{acknowledgments}


\end{document}